\documentclass[sn-mathphys,Numbered]{sn-jnl}

\usepackage{graphicx}%
\usepackage{multirow}%
\usepackage{amsmath,amssymb,amsfonts}%
\usepackage{amsthm}%
\usepackage{mathrsfs}%
\usepackage[title]{appendix}%
\usepackage{xcolor}%
\usepackage{textcomp}%
\usepackage{manyfoot}%
\usepackage{booktabs}%
\usepackage{algorithm}%
\usepackage{algorithmicx}%
\usepackage{algpseudocode}%
\usepackage{listings}%

\usepackage{amssymb, bm}
\usepackage{amsmath}
\usepackage{mathtools}
\newcommand{\angstrom}{\mbox{\normalfont\AA}}

\theoremstyle{thmstyleone}%
\theoremstyle{thmstyletwo}%

\theoremstyle{thmstylethree}%

\begin{document}

\title[Article Title]{Exploring Model Complexity in Machine Learned Potentials for Simulated Properties}


\author*[1]{\fnm{A.} \sur{Rohskopf}}\email{adrohsk@sandia.gov}

\author[1]{\fnm{J.} \sur{Goff}}

\author[2]{\fnm{D.} \sur{Sema}}

\author[2]{\fnm{K.} \sur{Gordiz}}

\author[3]{\fnm{N.C.} \sur{Nguyen}}

\author[2]{\fnm{A.} \sur{Henry}}

\author[1]{\fnm{A. P.} \sur{Thompson}}

\author[1]{\fnm{M. A.} \sur{Wood}}

\affil*[1]{\orgdiv{Center for Computing Research}, \orgname{Sandia National Laboratories}, \orgaddress{\city{Albuquerque}, \postcode{87185}, \state{NM}, \country{USA}}}

\affil[2]{\orgdiv{Department of Mechanical Engineering}, \orgname{Massachusetts Institute of Technology}, \orgaddress{\city{Cambridge}, \postcode{02139}, \state{MA}, \country{USA}}}

\affil[3]{\orgdiv{Department of Aeronautics and Astronautics}, \orgname{Massachusetts Institute of Technology}, \orgaddress{\city{Cambridge}, \postcode{02139}, \state{MA}, \country{USA}}}


\abstract{Machine learning (ML) enables the development of interatomic potentials that promise the accuracy of first principles methods while retaining the low cost and parallel efficiency of empirical potentials. 
While ML potentials traditionally use atom-centered descriptors as inputs, different models such as linear regression and neural networks can map these descriptors to atomic energies and forces.
This begs the question: what is the improvement in accuracy due to model complexity irrespective of choice of descriptors?
We curate three datasets to investigate this question in terms of \textit{ab initio} energy and force errors: (1) solid and liquid silicon, (2) gallium nitride, and (3) the superionic conductor LGPS.
We further investigate how these errors affect simulated properties with these models and verify if the improvement in fitting errors corresponds to measurable improvement in property prediction.
Since linear and nonlinear regression models have different advantages and disadvantages, the results presented herein help researchers choose models for their particular application.
By assessing different models, we observe correlations between fitting quantity (e.g. atomic force) error and simulated property error with respect to \textit{ab initio} values.
Such observations can be repeated by other researchers to determine the level of accuracy, and hence model complexity, needed for their particular systems of interest.}

\maketitle

\section{Introduction}\label{sec1}

Numerous macroscopic physical and chemical properties are predicted solely by the underlying interaction and motion of atoms.
Molecular dynamics (MD) simulations are an indispensable tool for studying this motion, and have enabled computational prediction and discovery in fields as diverse as materials science~\cite{choudhary2022}, energy storage~\cite{tafrishi2022, yao2022}, catalysis~\cite{vandermause2022}, and molecular biology~\cite{bai2022}.
High fidelity atomic forces for integrating Newton's equations of motion can be obtained from quantum mechanical calculations such as density functional theory (DFT), but unfavorable computational scaling of \textit{ab initio} methods prevents simulations of relevant length and time scales for many applications.
Ultimately, the accuracy of an MD simulation comes down to how these atomic forces are generated. 
To overcome this limitation, much progress has been made in the past decade by machine learning the quantum mechanical potential energy surface (PES)~\cite{deringer2019machine}.
These ML surrogates of the PES possess a favorable linear scaling with number of atoms like traditional empirical potentials, and have been shown to be more accurate in a number of scenarios~\cite{mishin2021machine}.
To achieve physically realistic and energy conserving simulations, ML potentials traditionally involve the use of an invariant atom-centered basis expansion that transform atomic environments into inputs suitable to ML models \cite{bartok2013representing}.
The mapping of these inputs to atomic energies and forces can take a variety of mathematical forms.
Common ML models span a range of functional complexity including linear regression, neural networks (NNs), or Gaussian processes.
Today a wide variety of descriptors (features, inputs, etc.) exist for atomistic machine learning, such as bispectrum components in the spectral neighbor analysis potential (SNAP)~\cite{thompson2015spectral}, complete bases such as atomic cluster expansion (ACE) descriptors~\cite{drautz2019atomic}, and traditional radial and angular bases such as in Behler-Parrinello atom-centered symmetry functions~\cite{behler2007generalized}.
Despite this wide development of descriptors for quantifying atomic environments, the effect of different ML models utilizing these descriptors as inputs remains relatively unexplored.

In this manuscript we quantify the gain in accuracy that can be expected when increasing model complexity, irrespective of choice of descriptors.
Here \text{model complexity} can be defined by both nonlinearity and number of fitting coefficients.
For example, a linear regression model with many fitting coefficients is more complex compared to a linear regression model with fewer fitting coefficients.
Likewise, a neural network model may be seen as more complex than linear regression models with a similar number of fitting coefficients.
Treating the \textit{models} and \textit{descriptors} of ML potentials separably gives insight where computational expense can be optimized in favor of accuracy. 
For completeness, we also consider three different descriptor sets that are available within LAMMPS; bispectrum components \cite{thompson2015spectral}, atomic cluster expansion \cite{drautz2019atomic}, and proper orthogonal descriptors \cite{nguyen2023proper}.

To compare the performance of different models irrespective of descriptor choice, we fit linear (which includes quadratic kernel tricks) and nonlinear (e.g. neural network) ML models with identical descriptors as input.
The gain in accuracy from using nonlinear models is quantified not only for validation errors in terms of \textit{fitting quantities} (energies and forces), but also how errors in \textit{fitting quantities} affect \textit{simulated properties}.
Here we focus on the distinction between \textit{fitting quantities}, which can include validation/test errors in atomistic quantities such as energies and forces, and \textit{simulated properties} which are calculated from using atomistic quantities like forces as inputs to calculations.
This distinction between categories of prediction accuracy is greatly needed in the field as the volume of published models has been rapidly increasing, and their translation to a larger MD user base warrants this detail of reporting.
To this end, we curate three datasets for comparing ML potential accuracy and its effect on select property predictions.
Additionally, exposing the accuracy limits that are determined by model form and/or descriptor basis uncovers physically interpretable insight into the nature of chemical bonds governed contained in the ground state PES to which these models are fit. 

First we benchmark the accuracy of linear, quadratic, and NN SNAP potentials on a silicon (Si) dataset containing solid and liquid phase \textit{ab initio} molecular dynamics (AIMD) configurations, and quantify the benefit a non-linear mapping from a single set of descriptors offers.
Next, we perform the same model comparisons with a more chemically complex gallium nitride (GaN) dataset containing AIMD trajectories from 300 K to 2300 K; here we observe if the accuracy advantage offered by NNs translates to improvements in phonon properties when using multiple descriptor sets (SNAP, ACE).
Finally, we curate a dataset for superionic diffusion in the solid electrolyte Li\textsubscript{10}Ge(PS\textsubscript{6})\textsubscript{2} (LGPS), to determine if complex model forms and descriptor sets offer an advantage over simple models for the macroscopically observable property of Li ion diffusion. 
We also benchmark our linear and NN models on another superionic conductor lithium phosphorus sulfide (LiPS) studied by graph neural networks (GNN) in literature \cite{park2021accurate}.

While many studies have shown model comparisons in \textit{ab initio} energy/force errors, it remains relatively unexplored how these errors correlate with the desired quantity of interest that the potential will be used for. 
Decoupling the effects of model form complexity from descriptor set is key in this understanding.
Fundamentally, we are attempting to resolve the accuracy gap between \textit{ab initio} and classical MD predictions, but for material properties that are sampled from a thermodynamic sampling of states.
To this end, we seek to elucidate the required accuracy in \textit{ab initio} energies/forces for accurately simulating material properties.
Since the definition of \textit{accurate} property prediction depends on individual research needs, it is important to show a range of model fitting errors and their associated simulated property errors.
Such knowledge will enable researchers to choose models that are accurate enough for their purposes, since more complicated but accurate models may introduce additional unnecessary cost to simulate and/or train.

\begin{figure}[htbp]
\centering
\includegraphics[scale=0.35]{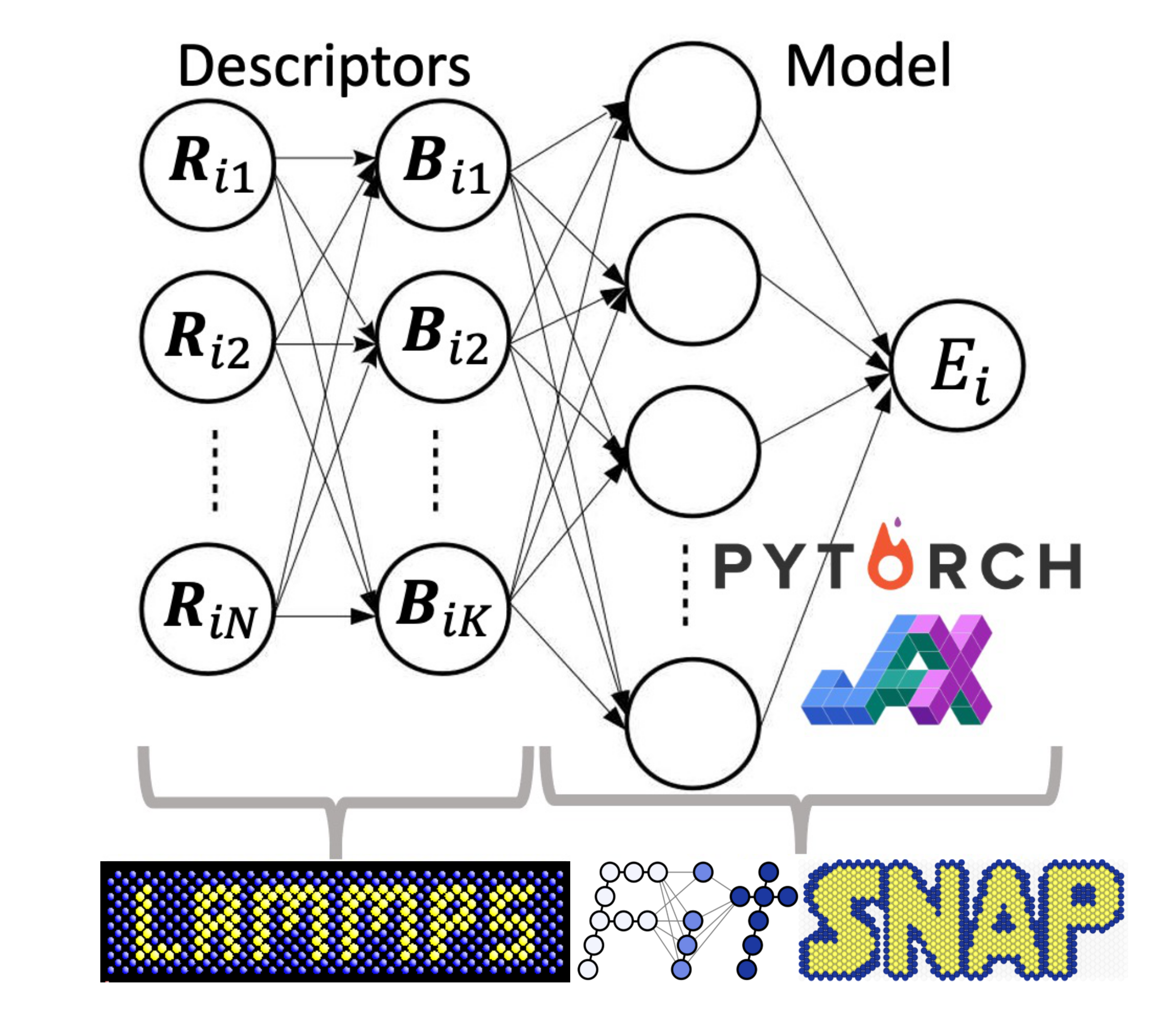}
\caption{Illustration of our distinction between ML potential descriptors and models. Interatomic distances $\bm R_{ij}$ between a central atom $i$ and neighbors $j$ are transformed into a set of $K$ descriptors denoted by a vector $\bm B_{i}$. In our work these features are calculated in LAMMPS and used as inputs for model optimization in FitSNAP, allowing seamless deployment to LAMMPS for MD simulations. Note that the treatment of atom types is not illustrated here, different model coefficients are prescribed for each element. Current LAMMPS supported descriptors included SNAP and ACE, which we can use in different models with FitSNAP. FitSNAP also provides a custom descriptor calculator that simply takes in a LAMMPS neighbor list, which is useful for prototyping different descriptors or using graph neural networks.}
\label{fig:diagram}	
\end{figure}

It is worth noting that these trade-offs are only exposed in ML interatomic potentials, where traditional empirical potentials (Tersoff, EAM, COMB, ReaxFF) have a fixed accuracy with respect to computational cost.
Decisions regarding increased cost of using complex NN models for the gain in accuracy, compared to simpler and more performant linear models are chosen by the user.
Linear models also come with a range of other advantages, such as fast training times and convenient measures of Bayesian uncertainties \cite{zhu2022fast}, so understanding the sacrifice in accuracy will help researchers weigh these pros and cons.
To facilitate researchers in using our results to choose and develop their own models, we utilize the open source FitSNAP software as a ML interface to LAMMPS, with capabilities to fit models of varying complexity while keeping descriptor settings constant.
FitSNAP allows training of various model/descriptor combinations, e.g. SNAP or ACE descriptors with both linear or neural network models.
Our distinction between the ML potential \textit{descriptors} and the \textit{model} is illustrated in Figure \ref{fig:diagram}.
Note that graph neural networks of atomic interactions will replace the mapping of $\{\bm R_{i1}, \ldots, \bm R_{iN} \}\rightarrow \bm B_{i}$ with additional interaction layers to construct an analogous descriptor.

The modular separation between \textit{descriptors} and \textit{models} in Figure \ref{fig:diagram} allows us to take advantage of performant descriptor and descriptor gradient calculations in LAMMPS, which are then input to automatic differentiation frameworks such as PyTorch \cite{paszke2019pytorch} and JAX \cite{bradbury2018jax} for optimization.
Combining the LAMMPS implementation of descriptors with ML frameworks such as PyTorch allows performant NN force training via a modified iterated backpropagation algorithm, explained in more detail in the Methods section.
This modular separation between descriptors and models also allows researchers to explore which descriptor/model combination best suit their needs.
While this separation of descriptor and model offers distinct advantages, it is important to note that this separation is entirely abstract and conceptual.
Indeed, NNs may be viewed as linear models if considering the final layer of the network, which can be understood as a transformed set of descriptors that combine linearly at the output.
In this view, the question becomes how much does a NN improve the original set of descriptors.
Nonetheless we seek to answer how much fitting accuracy is improved by feeding descriptors into different models, and quantify the improvement in simulated property prediction therefrom.
We begin by considering analytical differences between some ML potential models.

\section{ML Potential Model Forms}

In general, atom-centered ML potentials seek a relation between energy of an atom and its environment.
The total energy is then a sum over all atomic energies, written as
\begin{equation}
\label{energy_general}
E = \sum_i E_i (\bm B_i)
\end{equation}
where $\bm B_i$ is the feature vector for atom $i$, otherwise known as the descriptors which quantify the atomic environment.
Linear ML potentials are the simplest models by assuming a linear relationship between descriptors and the atomic energy, so that
\begin{align}\label{ei_linear}
\begin{split}
E_i = & \ \beta_0 + \sum_k^K \beta_{k} B_{ik} \\
    = & \ \beta_0 + \bm \beta \cdot \bm B_i
\end{split}
\end{align}
where $\bm \beta$ is a vector of $\beta_{k}$ model coefficients corresponding to descriptor $k$ out of $K$ descriptors.
For multi-element systems, we may use a different set of coefficients for each element type as in weighted density descriptors \cite{thompson2015spectral}, or a unique set of coefficients for combinations of elements as in ACE \cite{drautz2019atomic} or chemSNAP \cite{cusentino2020explicit}.
Nonlinear terms, hence more complexity, may be introduced in linear models via kernel tricks; one popular example here involves quadratic ML potentials such as quadratic SNAP \cite{wood2018extending}.
Here, one may think of the energy in Eq. \ref{ei_linear} as a first order Taylor expansion in descriptor space.
A second order expansion may be written as
\begin{equation}
\label{ei_quad}
E_i = \beta_0 + \bm \beta \cdot \bm B_i + \frac{1}{2} (\bm B_i)^T \cdot \bm \alpha \cdot \bm B_i
\end{equation}
where $\bm \alpha$ is a symmetric $K \times K$ matrix of model coefficients.
The quadratic potential of Eq. \ref{ei_quad} is readily optimized in the same linear regression manner as Eq. \ref{ei_linear}, with the cost of significantly more fitting coefficients.
This increased number of coefficients, however, offers more flexibility and improved fitting errors \cite{wood2018extending, zuo2020performance}.
Even with their improvements in accuracy, quadratic models can retain computational costs close to linear models since we may use the same descriptors for the first and second order terms as seen in Eq. \ref{ei_quad}.
Of course quadratic models are more expensive to train due to their greatly increased number of coefficients, but this training cost can still be cheaper than more complicated nonlinear models such as neural networks.
Atom-centered neural network potentials approximate atomic energies with multilayer perceptrons that are functions of the descriptors, written in terms of matrix products of layer weights as
\begin{equation}
\label{ei_nn}
E_i = \bm W_L \sigma(...\sigma(\bm W_2 \sigma(\bm W_1 \bm B_i)))
\end{equation}
where $\bm W_l$ denotes the weights for layer $l$, and $\sigma$ is the activation function used at each layer.
The last node of the multilayer perceptron predicts a scalar energy $E_i$, and is therefore not transformed with the activation function $\sigma$.
One may expect that NN models such as Equation \ref{ei_nn} offer more complexity/flexibility than the linear and quadratic models in Equations \ref{ei_linear} and \ref{ei_quad} since there is no limit on the degree of nonlinearity in NN models, and because of the universal approximation theorems \cite{hornik1989multilayer, csaji2001approximation}.
Whether or not the improved accuracy offered by this extra flexibility matters in representing the true PES or for modelling material properties, however, is an unexplored question that we investigate in this study.

We may better appreciate how these model types are related and how they offer different degrees of flexibility for modelling the PES by considering a general view of interatomic forces for atom-centered ML potentials.
The force on atom $i$ arises from the negative gradient of the total potential energy in Equation \ref{energy_general} with respect to the position of atom $i$.
For atom-centered ML potentials, these forces are generally written as
\begin{equation}
\label{force_general}
\bm F_i = - \sum_j \sum_k \beta_{jk} \frac{\partial B_{jk}}{\partial \bm R_i}
\end{equation}
where $\beta_{jk}$ represents the derivative $\partial E_j / \partial B_{jk}$ for each neighboring atom $j$.
All the aforementioned models possess different $\beta$ that offer varying degree of flexibility when modelling interatomic forces, shown mathematically below.
\begin{equation}\label{betas}
  \beta_{jk} = 
  \begin{dcases*} 
  \text{$\beta_k$} & (Linear) \\ 
  \text{$\beta_k + \sum_l \alpha_{kl} B_{jl}$} & (Quadratic) \\
  \text{$\frac{\partial}{\partial B_{jk}} \{\bm W_L \sigma(...\sigma(\bm W_2 \sigma(\bm W_1 \bm B_j))) \}$} & (NN)
  \end{dcases*} 
\end{equation}
Equation \ref{betas} shows that linear ML potentials assume neighboring force contributions $F_{ij}$ are proportional to a the descriptor derivatives by the coefficient $\beta_k$.
Meanwhile, $\beta_{jk}$ for quadratic models is a linear combination of all other descriptors.
Finally, NNs provide the highest possible order of nonlinearity since the derivatives of a multilayer perceptron with respect to the inputs is yet another multilayer perceptron.
In principle Equation \ref{betas} shows that NNs offer the most flexibility in modelling interatomic forces, by not assuming any functional form of the descriptors.
It remains to be seen, however, how much this extra complexity affects force accuracy, and whether this improvement in force accuracy improves property prediction. 
To begin investigating this question we first apply these models to silicon, one of the simplest systems benchmarked by ML potentials. 

\section{Solid and Liquid Silicon}

In the literature, a Si dataset was used to demonstrate state-of-the-art accuracy with the recently developed ACE descriptors \cite{lysogorskiy2021performant}, and another dataset was developed to demonstrate the ability of GAP to model a wide variety of phase spaces and properties \cite{bartok2018machine}.
To this end, silicon has become a common benchmark system for simple comparisons.
Here, we curate a simple dataset for the purpose of comparing different models on phonon properties and liquid structure.
Our dataset includes AIMD configurations from 300 K to the melting point, along with 2000 K liquid AIMD simulations, with more details in the SI.
Before simulating properties with our potentials, however, we seek to benchmark their accuracy on the energies and forces within the dataset.
This will show to what extent more accurate fits give better \textit{ab initio} property agreement.
In this regard, we seek a fair method of comparing different model accuracies on fitting quantities irrespective of training hyperparameters, which requires a closer look at the loss function.

Hyperparameters common to all ML potential models and optimization methods reside in the loss function.
When training to energies and forces we use the same L2 norm loss function for all models, given by
\begin{equation}
\label{loss_function}
L = \frac{1}{M}\sum_m^M \frac{1}{N_m} \left[ w_m^E \left( \hat{E}_m - E_m \right)^2 + \frac{w_m^F}{3} \sum_{a,i}^{3N_m} \left( \hat{F}_{a,i,m} - F_{a,i,m} \right)^2 \right]
\end{equation}
where $M$ is the number of configurations, $N_m$ is the number of atoms in configuration $m$, $\hat{E}_m$ is the model energy of a configuration, and $E_m$ is the target \textit{ab initio} energy.
The right-most term is a L2 norm in force components where $\hat{F}_{a,i,m}$ is the model force component on atom $i$ in configuration $m$ with Cartesian direction $a$.
Likewise for the target \textit{ab initio} force components $F_{a,i,m}$.
For linear and NN models we solve this L2 norm via singular value decomposition (SVD) and gradient descent, respectively, with more details in the SI.
Regularization penalty functions were not used, but are available within the FitSNAP framework.
Quadratic models are also readily solved via SVD since they can be written as a linear combination of descriptors as in Equation \ref{ei_quad}.
Linear models can trained with gradient descent, but SVD offers fast training times that have been utilized to create successful potentials in the past \cite{thompson2015spectral, wood2019data, cusentino2020explicit, zuo2020performance, nikolov2021data, nguyen2021billion}, so we wish to retain this advantage in a model comparison.
The common hyperparameters between linear and nonlinear models are therefore loss function energy weights $w_m^E$ and force weights $w_m^F$.

\begin{figure}[htbp]
\centering
\includegraphics[width=\textwidth]{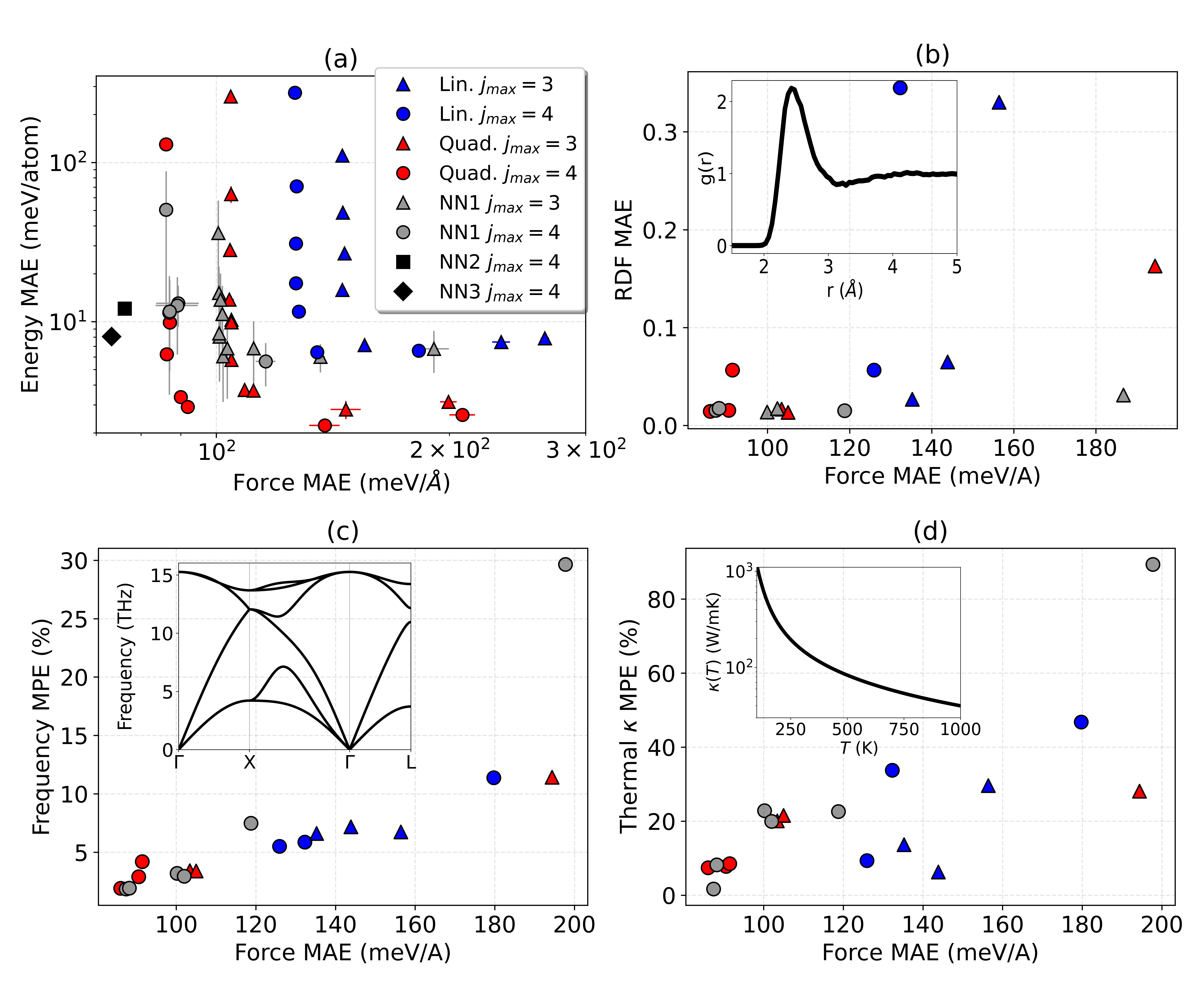}
\caption{(a) Force/energy tradeoff for linear, quadratic, and NN models with SNAP descriptors trained on our silicon data set. Different SNAP descriptor settings are $j_{max} = 3$ (31 descriptors) and $j_{max} = 4$ (56 descriptors). Each point represents a different force/energy weight ratio. The reported errors are an average of 5 random validation set errors with horizontal and vertical lines showing standard deviation in energy and force, respectively. NN1 refers to a 32 $\times$ 32 hidden node architecture, NN2 is 250 $\times$ 250, and NN3 is 500 $\times$ 2000. (b) Radial distribution function mean absolute error over the radii shown in the inset, between potentials and AIMD. (c) Phonon frequency mean percent error across the Brillouin zone shown in the inset. (d) Thermal conductivity mean percent error across the range of temperatures from 100 K to 1000 K, as shown in the inset. Note that more complex models (quadratic and NN) possess the lowest force errors and accurately capture all simulated properties.}
\label{fig:fig1}	
\end{figure}

For linear models, effort is often dedicated to optimizing these energy/force weight hyperparameters by looping over SVD fits combined with multi-objective optimization on the energy/force weights \cite{thompson2015spectral, wood2019data, cusentino2020explicit, zuo2020performance, nikolov2021data, sikorski2022machine}.
Gradient descent training for NNs is more costly, however, so we do not individually optimize the weight hyperparameters here.
Instead, we perform fits with a variety of force/energy weight ratios $w^F/w^E$, where the same ratio is applied to all configurations in the training set.
These $w^F/w^E$ ratios are chosen in a range of $10^{-4}$ to $10^{4}$, with more details on the exact values in the SI.
By comparing the best energy and force errors on a validation set using a variety of $w^F/w^E$ ratios, we achieve a fair comparison of model accuracy irrespective of loss function hyperparameters.
For an initial comparison we fit linear, quadratic, and NN models on our silicon data set using only SNAP descriptors, to observe the effect of model complexity alone.
Descriptor complexity is also observed by using SNAP descriptor settings $j_{max} = 3$ (31 descriptors) and $j_{max} = 4$ (56 descriptors).
We refer the reader to literature for understanding how the $j_{max}$ parameter influences the number of SNAP descriptors \cite{thompson2015spectral}.
For all $w^F/w^E$ ratios, we report the average validation error on five random 10\% validation sets, along with a standard deviation thereof.
Scanning a variety of $w^F/w^E$ ratios we find that all models saturate at different energy and force accuracy, as seen in Figure \ref{fig:fig1}a, with notable trade offs between these two fitted properties.

It is important to note the force error saturation experienced by all models as shown in Figure \ref{fig:fig1}.
We proposed this saturation level should be used to fairly compare how well different models can match the PES shape or its derivatives.
Furthermore, properties that rely on sampling a MD trajectory at equilibrium are hypothesized to be highly sensitive to force accuracy that determines atomic motions in some equilibrium state.
Other immediately observable and important trends involve the improvement in force accuracy due to increasing model complexity, irrespective of descriptor choice.
For example, SNAP descriptors with $j_{max} = 3$ saturate near ~150 meV/$\angstrom$, as seen with the blue triangles in Figure \ref{fig:fig1}.
Simply feeding these same descriptors into quadratic or small NNs ($32 \times 32$ hidden node architecture denoted by NN1 in Figure \ref{fig:fig1}) decreases the force saturation level down to ~100 meV/$\angstrom$, as seen with the red and grey triangles.
A similar improvement is seen with more detailed SNAP descriptors using $j_{max} = 4$, where the linear models (blue circles) saturate at ~125 meV/$\angstrom$ while quadratic and small NN models (red and grey circles, respectively) saturate at ~80 meV/$\angstrom$.
Further complicating the model with larger NNs, however, does not offer much improvement in forces.
This is shown with the black diamond and square in Figure \ref{fig:fig1}a, representing NNs with architectures of $250 \times 250$ and $500 \times 2000$ node architectures, respectively.
Using these much larger models only yielded a ~5 meV/$\angstrom$ improvement over the smaller NNs.
This agrees with previous experiments in literature, which found deeper/wider NNs have little effect on energy accuracy \cite{montes2022training}.
Nonetheless we observe a clear improvement in force accuracy due to quadratic and NN models over linear models, using the same descriptors.
This begs the question: does this improvement in forces translate into accurate dynamics and material properties that require sampling a trajectory?

Although it has been noted in literature that force errors in general are not enough to say a potential is reliable for stable MD or property prediction \cite{fu2022forces}, we find for our models here that force error correlates well with liquid structure and phonon property error.
The easiest dynamical quantity to reproduce was the 2000 K liquid RDF in Figure \ref{fig:fig1}b, where we achieved low RDF MAE ($<$ 0.01 RDF units) with models possessing up to $\sim$ 120 meV/$\angstrom$ force error.
More liquid structure data such as angular distribution functions for specific $w^F/w^E$ ratios are shown in the SI.
Solid state phonon frequencies, on the other hand, exhibited a pronounced correlation with force error as shown in Figure \ref{fig:fig1}c. 
This is not surprising due to the fact that vibrational frequencies depend on second order derivatives of the PES, so models that best match forces (first derivative) should better capture how force changes with atomic displacement.
Indeed, the only potentials capable of reaching 1\% frequency error are those with the lowest force errors; quadratic and NN SNAP with $j_{max} = 4$.
Likewise for thermal conductivity error in Figure \ref{fig:fig1}d, the same quadratic and NN SNAP $j_{max} = 4$ potentials accurately reproduced thermal conductivity within 10\%.
These quadratic and NN SNAP $j_{max} = 4$ potentials achieved the lowest force errors of 80 meV/$\angstrom$ and were the only potentials of simultaneously reproducing all AIMD quantities (liquid structure, phonon frequencies, and thermal conductivity) to within reasonable agreement (0.01 RDF MAE, 1\% frequency error, and less than 10\% thermal conductivity error).
This shows the practical advantage of using atom-centered descriptors such as SNAP with more complex models, and that the 50 meV/$\angstrom$ improvement in forces can yield a noticeable improvement in solid/liquid quantities of interest.
It is important to note that this does not come with a significant increase in computational cost when simulating MD, as most of the cost is associated with the descriptor calculation.
More details on computational cost are reported in the SI, and be aware that software factors such as PyTorch interfaces to allow NN potentials in LAMMPS are constantly changing.
Aside from the slightly increased computational cost of MD, researchers should also acknowledge the increased cost of training for NN models.

The SVD procedure for training linear SNAP on a data set of ~2000 configurations like we use here completes on the order of minutes.
Meanwhile, the 1000 epochs of gradient descent training used for our NNs can take 12 hours on a single CPU core with our silicon data set.
Linear solvers therefore possess a notable advantage of fast training times, allowing one optimize other hyperparameters like those associated with SNAP descriptors to match more important quantities, such as stability metrics, defect/surface energies, and others \cite{sikorski2022machine}.
Noting these advantages in training cost, one may be attracted to the prospect of using quadratic models, especially since they saturate near the same force errors as compared to NNs.
Both quadratic and small NN models with different SNAP descriptors $j_{max} = 3$ and $j_{max} = 4$ converge at similar force errors near 100 meV/$\angstrom$ and 80 meV/$\angstrom$, respectively.
This might suggest that quadratic and NN models arrive at the same approximation of the \textit{ab initio} PES, which is important to know because researchers should not waste time fitting both models if they are indeed the same solution.

\begin{figure}[htbp]
\centering
\includegraphics[width=\textwidth]{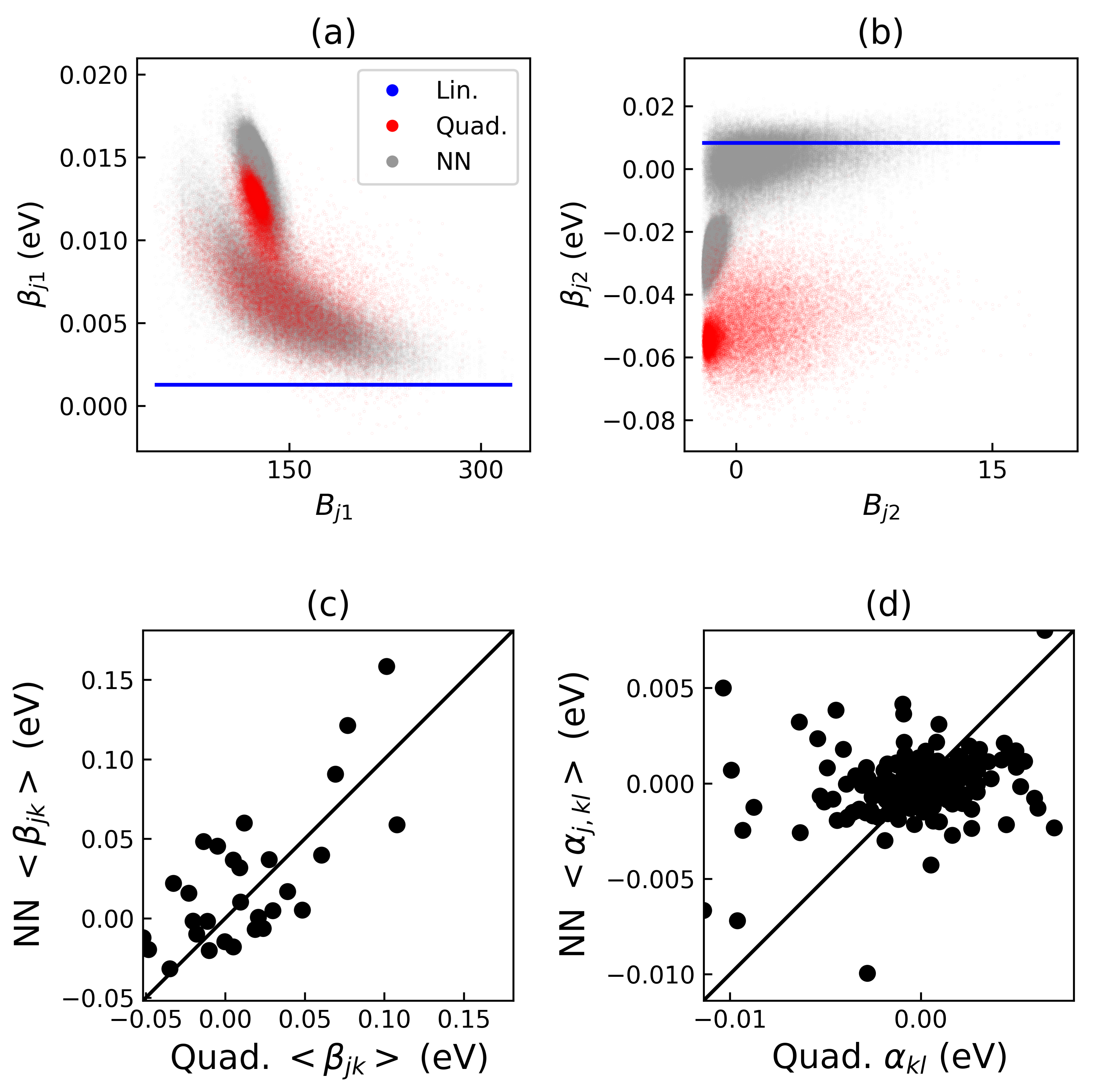}
\caption{Comparisons of model derivatives with respect to descriptors for our silicon data set, showing that quadratic and small NNs arrive at similar solutions of the PES up to first order in descriptors. (a) First derivatives of energy with respect to the first bispectrum component, denoted by $\beta_{j1}$, as a function of the first bispectrum component $B_{j1}$ for each atom $j$. These first derivatives are constants for a linear model, but unique for each atom in the case of quadratic and NN. For this first bispectrum component, there is significant overlap between quadratic and NN. (b) First derivative of energy with respect to the second bispectrum component, showing less overlap. (c) Parity plot showing a correlation between quadratic and NN $<\beta_{jk}>$ averaged over all atoms $j$ in the data set. (d) Parity plot for second derivatives $<\alpha_{j, kl}>$ averaged over all atoms, which is constant for quadratic models. }
\label{fig2}	
\end{figure}

To test whether quadratic and NN models arrive at the same solution, we compare the derivatives of the PES with respect to model descriptors.
These first derivatives are the $\beta_{jk}$ given in Equation \ref{betas}, and plotted as a function of their respective bispectrum components in Figure \ref{fig2}.
By comparing model first derivatives with respect to descriptors, we are comparing the shape of the PES up to first order in the descriptors.
Some bispectrum components share similar first derivatives between quadratic and NN SNAP models, e.g. Figure \ref{fig2}a shows that the clouds of first derivatives of the first bispectrum components significantly overlap for all data points when evaluated by quadratic and NN models trained on the silicon set.
The derivative with respect to the second bispectrum component shown in Figure \ref{fig2}b, however, shows less overlap.
Overall, we find a correlation in the average value of first derivatives with respect to descriptors, shown by the parity plot for quadratic and NN $\beta_{jk}$ shown in Figure \ref{fig2}c.
Quadratic and NN models therefore arrive at similar approximations of the PES up to first order in the descriptors.
Up to second order, however, the quadratic and NN models are different, shown by the parity plot in average second derivatives $\alpha_{j,kl} = \partial^2 E_j / \partial B_{jk} \partial B_{jl}$.

Although quadratic and NN models arrive at different approximations of the PES up to second order, these differences do not significantly affect force, phonon property, or liquid structure accuracy.
This may be surprising since quantities like vibrational frequencies directly depend on spatial force derivatives; at some level, properties become insensitive to differences in PES approximations, and we seemed to have reached this point with quadratic and NN SNAP.
Though we acknowledge there is a deeper study to be had of which material properties would be sensitive to these differences, we highlight the unique solutions provided by these two model forms.
It therefore might be advantageous for researchers to fit both models and retain all fits even though they have similar errors, because differences up to second or higher orders may be responsible for other phenomena not studied here, such as improvements in stability or extrapolation to much different phase spaces.
Exploring the differences in accuracy for other systems will also show whether the agreement between quadratic and NN models is consistent.
Since silicon is a relatively simple single element system, we can introduce training complexity by considering a system with two element types.
To that end, we see if similar model trends arise by considering our gallium nitride AIMD data set.

\section{Gallium Nitride Thermal Transport}

GaN is an important material in power electronics where large heat fluxes limit performance in devices \cite{tsao2018ultrawide}.
Phonon transport in GaN is therefore well studied and there are reports in literature of accurate potentials for GaN \cite{minamitani2019simulating}.
Here we quantify the improvement of quadratic and NN models over linear models like we did for Si.
We perform this comparison using descriptors that treat multi-element systems to see if NNs offer noticeable gain in accuracy with different types of descriptors.
Specifically, we now add ACE descriptors to the set of comparisons due to their explicit treatment of multi-element pairs \cite{drautz2019atomic}.
In the ACE formalism each pair of atoms has its own set of descriptors and model coefficients based on element type.
For example, GaN has 4 sets of coefficients encompassing Ga-Ga, Ga-N, N-Ga, and N-N interactions.
Meanwhile, traditional "weighted density" SNAP reduces all element types in the environment onto a single descriptor value, with separate model coefficients reserved only for distinct elements as the central atom instead of element-element pairs.
There is also an explicit multi-element version of SNAP, but we have not introduced this formalism with NNs within FitSNAP \cite{cusentino2020explicit}.

\begin{figure}[htbp]
\centering
\includegraphics[width=\textwidth]{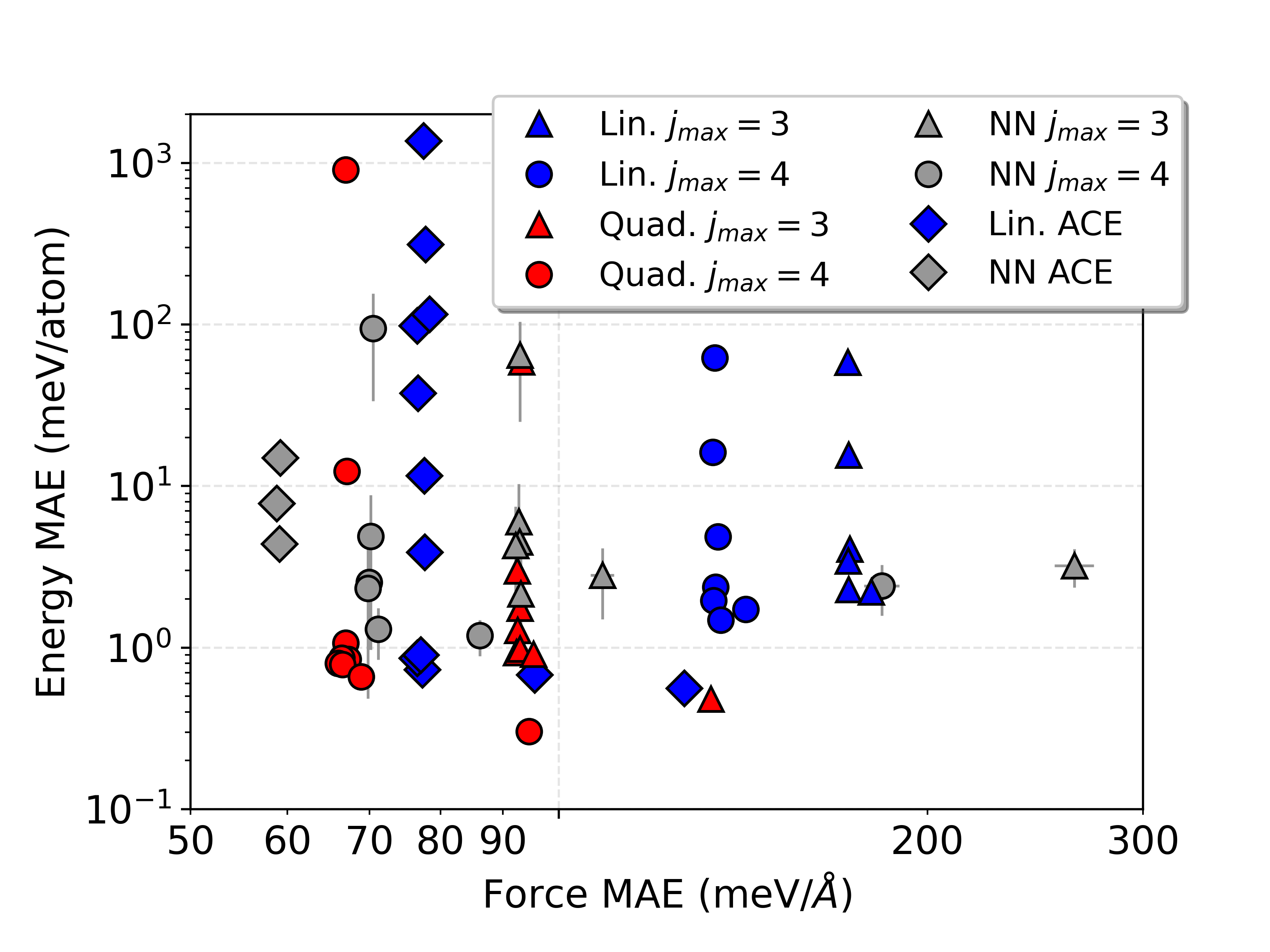}
\caption{GaN force and energy saturation for different models using SNAP and ACE descriptors. Each point represents a model fit with a different force/energy ratio. Blue symbols are linear models with different descriptors, red symbols are quadratic models, and grey symbols are NN models. Note the same force saturation behavior as seen for Si, with more flexible models saturating at lower force errors. It is important to note that our choice of ACE descriptors was limited by the amount we could store in RAM for NN models, since we store all the descriptor derivatives in Equation \ref{force_general} in RAM with the current FitSNAP implementation. This could be circumvented by using autograd all the way back to atomic positions for calculating the force, instead of storing descriptor gradients in memory. It is important to note that ACE is an extremely flexible potential and one could construct a more accurate ACE potential than presented here by either adding more basis functions and/or increasing body order; for our purposes we simply explore the benefit of using NNs with a constant set of descriptors.}
\label{fig3}	
\end{figure}

Due to explicit multi-element treatment it is therefore expected that ACE can achieve lower errors compared to weighted density SNAP; this has been noted recently when comparing ACE with SNAP and many other potentials \cite{lysogorskiy2021performant}.
When feeding ACE and SNAP descriptors into NNs, however, we use the same multi-element treatment of atom-centered NN models.
Here we assign a unique NN to each central atom based on its type, as common in literature for ANI \cite{gao2020torchani} and Behler-Parrinello NNs \cite{behler2007generalized}.
The force/energy tradeoff for GaN using SNAP descriptors with various models, and ACE descriptors with linear and NN models, is shown in Figure \ref{fig3}.
Linear ACE saturates at a lower force error than linear SNAP.
This is not surprising, especially since the ACE settings we used result in 120 descriptors for each element type pair, while SNAP $j_{max} = 3$ and $j_{max} = 4$ involve only 31 and 56 descriptors for each element, respectively.
Our ACE basis could therefore be thought of as more comprehensive than our SNAP basis used here.
One may therefore expect less of an improvement in errors when considering a nonlinear ACE model, since the descriptors already offer a comprehensive description of the environment.
Indeed, NN ACE only exhibits an improvement in forces from ~80 meV/$\angstrom$ to ~60 meV/$\angstrom$ when keeping descriptors constant.
This improvement is also limited by our implementation of multi-element NNs; linear ACE possesses unique coefficients for each element type pair but our NNs are unique for each element type.
It would be more commensurate to use different NNs for each element type pair, which should offer more flexibility.
Nonetheless we see less of an improvement compared to SNAP descriptors, which experience almost a two-fold reduction in force errors as seen in Figure \ref{fig3}.

\begin{figure}[htbp]
\centering
\includegraphics[width=\textwidth]{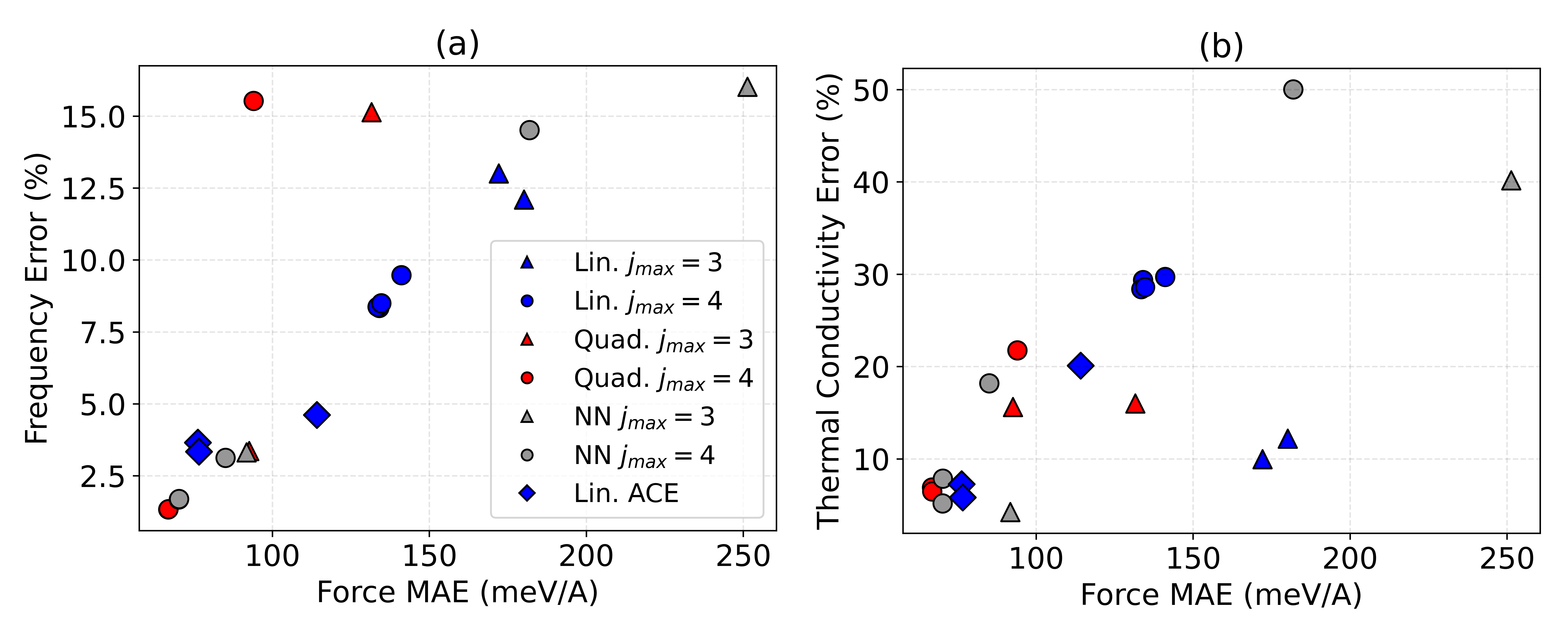}
\caption{(a) Correlation between force error and phonon frequency error for various potentials. Each point represents a fit with a different force/energy weight ratio. (b) Correlation between force error and thermal conductivity error. Note that thermal conductivity is a higher order property and therefore some potentials may obtain small errors for the wrong reasons, so there is less of a correlation compared to frequencies. It is important to note that with a sufficient basis, such as ACE, one can achieve low property errors with linear models on par with nonlinear models using other bases. Here we used the same linear ACE model from Figure \ref{fig3}, although more comprehensive ACE bases could be used as well.
TODO: Add a more comprehensive ACE basis that achieved ~50 meV/A.
}
\label{fig4}
\end{figure}

With SNAP descriptors, we see the same saturation of quadratic and NN models at similar force errors, as we saw in Si. 
This improvement in forces has a measurable effect on vibrational frequencies and thermal transport, as shown in Figure \ref{fig4}.
In Figure \ref{fig4} we see a clear positive correlation between force error and phonon frequency error averaged over symmetry directions in the Brillouin zone, with more details in the SI.
In Figure \ref{fig4} we observe a weaker correlation between force error and thermal conductivity error.
This is expected since thermal conductivity is a higher order property and therefore some potentials may obtain small errors for physically improper reasons, such as cancellation of errors due to improper phonon contributions. 
There is therefore less of a correlation compared to frequencies, as we also saw for Si.
Nonetheless, quadratic and NN models with $j_{max} = 4$ offer enough flexibility to simultaneously achieve low frequency errors of 1\% and thermal conductivity errors $<$ 10\%.
The two-fold improvement in forces when using SNAP descriptors with quadratic and NN models is therefore important for describing phonon transport in GaN.
We note that past efforts in modelling GaN involved the use of Taylor expansion potentials whose fitting coefficients are the PES spatial derivatives, which allowed nearly exact reproduction of \textit{ab initio} phonon dispersion  \cite{rohskopf2020fast}.
The potentials made here may not be as computationally efficient as low order Taylor expansions but retain the ability to be used in scenarios beyond the solid phase that Taylor expansions are limited to, such as melting or diffusion.

\section{Lithium Ion Diffusivity}

Si and  GaN are still relatively simple systems from a molecular modelling standpoint and therefore served as good candidates for observing improvements in accuracy, without confounding variables such as complicated chemical forces or other difficult training data.
These systems alone, however, may not be as commensurate with other efforts of the community to study systems with more element types and more complicated dynamical properties.
We therefore seek evermore difficult multi-element systems and dynamical properties to benchmark our linear and nonlinear models on.
For this purpose we turn to Li ion conductors, which were recently used in literature to exhibit profound ability to match AIMD properties with state-of-the-art graph neural network potentials \cite{batzner20223}.

Li ion superionic conductors are solids that exhibit Li diffusivity on par with liquid electrolytes \cite{boyce1979superionic}.
These materials are of recent interest in the energy storage community because they may contribute to the commercialization of all-solid-state batteries \cite{kato2020li10gep2s12}.
It is therefore of interest for researchers to study the atomistic mechanisms allowing for superionic diffusion, which may allow engineering of materials with favorable diffusion or synthesis properties \cite{he2019crystal}.
To achieve this level of atomistic insight, interatomic potentials for Li superionic conductors are required.
ML potentials for Li ion systems to date use deep neural networks that resulted in excellent agreement with AIMD diffusion \cite{winter2022simulations, batzner20223}.
These systems remain somewhat challenging for the ML potential community, with only a few NN potentials available.
It may therefore be believed that NN models are required to model such systems with 3-4 element types and unique diffusion mechanisms.
This model complexity of existing Li ion potentials therefore suggests that energies and forces cannot be represented by low order functional forms of atom-centered descriptors.

To test whether NNs offer advantages for modelling Li superionic conductors as we saw with Si and GaN, we curate an AIMD data set for LGPS.
This dataset contains AIMD trajectories from 300 K to the melting point of around 2000 K, with more details in the SI.
We trained a variety of potentials to 5,000 AIMD configurations sampled across these temperatures.
From our experience with Si and GaN in Figures \ref{fig:fig1} and \ref{fig3}, we compare force saturation errors since this is a fair comparison for how well the models can fit the PES.
Force errors associated with our best fits are tabulated in Table \ref{table:1}.
Validation will be performed by simulating ion diffusion at 600 K.
\begin{table}[h!]
\centering
\begin{tabular}{||c c c c c||} 
 \hline
 Lin. SNAP1 & NN SNAP2 & NN SNAP1 & POD & POD+SNAP \\ [0.5ex] 
 \hline\hline
 180.4 & 116.2 & 84.5 & 70.4 & 48.1 \\ 
 \hline
\end{tabular}
\caption{Force MAE (meV/$\angstrom$) of various potentials for LGPS. SNAP1 refers to $j_{max} = 3$ (31 descriptors) and SNAP2 refers to $j_{max} = 4$ (56 descriptors). For POD+SNAP, we use $j_{max} = 2$ which includes 15 4-body SNAP descriptors on top of the 91 2-body and 3-body POD descriptors.}
\label{table:1}
\end{table}
Here we use linear SNAP with $j_{max} = 4$ and NN SNAP with $j_{max} = 3$ and $j_{max} = 4$.

To benchmark more linear models on Li conductors, we include another recently developed potential based on proper orthogonal decomposition \cite{nguyen2023proper}.
POD descriptors use a form of proper orthogonal decomposition traditionally applied in continuum mechanics, formulated for discrete atomistic applications.
Our POD potential here has a total of 3314 coefficients composed from 11 2-body descriptors and 80 3-body descriptors.
These POD descriptors treat multi-element interactions such that each atom pair or triplet has its own coefficients, but use the same basis functions, so that the computational cost is independent of the number of elements.
The number of fitting coefficients for our linear SNAP model is number of descriptors times number of elements resulting in $56 \times 4 = 224$ fitting coefficients.
It is therefore not surprising that linear POD alone obtains much better force errors than linear SNAP.
As has been seen before with Si and GaN, simply adding more descriptors to a model results in a notable increase in accuracy.
Adding SNAP descriptors to POD for LGPS also reduced force error from 63.3 meV/$\angstrom$ to 48.1 meV/$\angstrom$.
Note there are diminishing returns on accuracy with increased basis size, as was mentioned in Wood et al. \cite{wood2018extending}.

\begin{figure}[htbp]
\centering
\includegraphics[width=\textwidth]{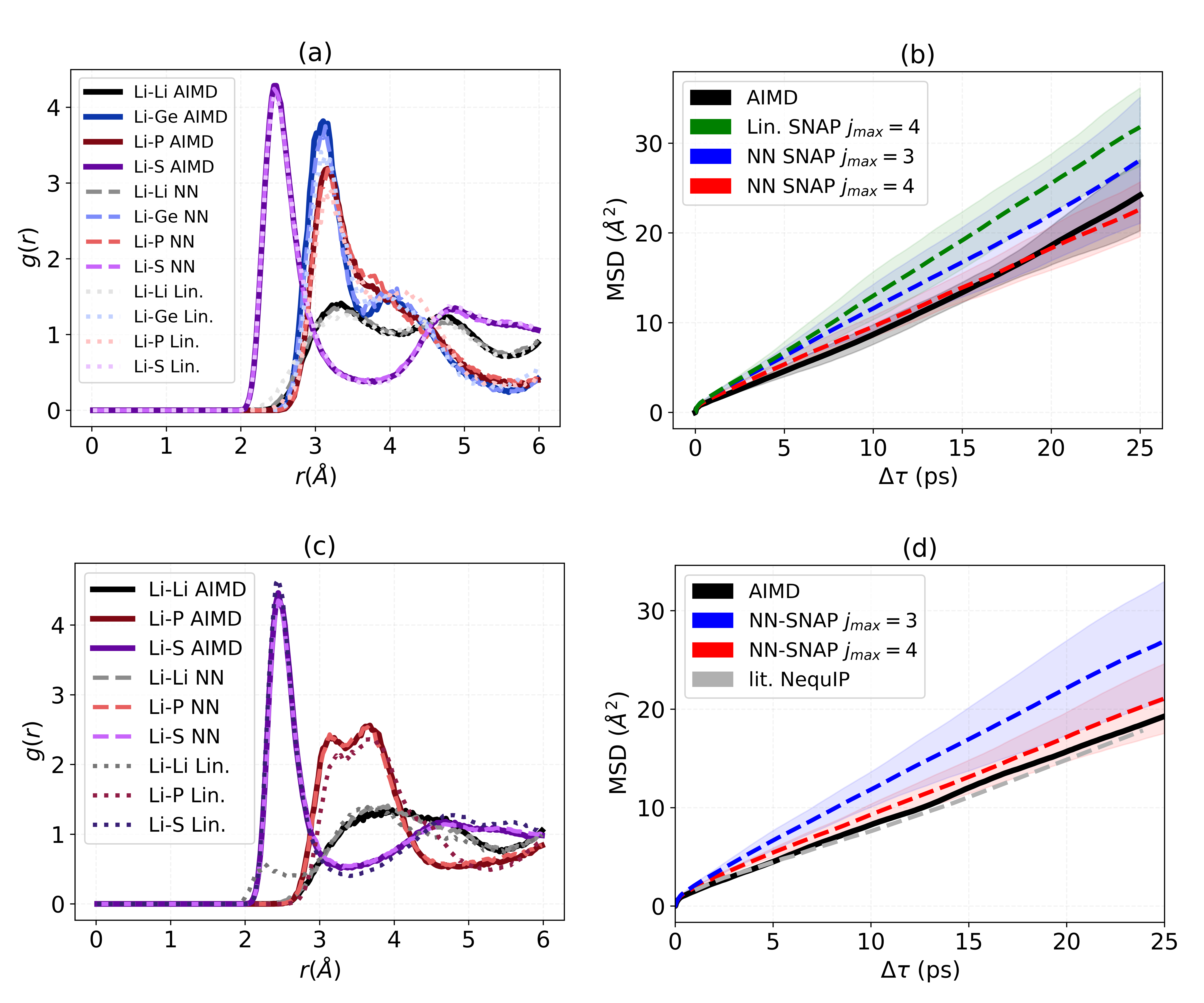}
\caption{LGPS and LiPS structure and diffusivity. (a) LGPS RDFs for Li ions using linear and NN SNAP potentials compared to AIMD at 600 K. (b) LGPS mean square displacement at 600 K for the SNAP potentials. (c) LiPS RDFs for Li ions using linear and NN SNAP potentials, showing the linear SNAP could not properly stabilize LiPS with the given literature trajectory. (d) LiPS mean square displacement at 520 K simulated with NN SNAP potentials and data taken from the NequIP GNN potential results from literature \cite{batzner20223}.}
\label{fig:lgps_lips}	
\end{figure}

To observe the effect of model complexity on simulating ion diffusion, we first focus on SNAP descriptors input to linear and NN models.
MD simulation results for our LGPS data set and a literature LiPS data set are shown in Figure \ref{fig:lgps_lips}.
From \ref{fig:lgps_lips} we see a clear improvement in MD simulations of ion diffusion using NN SNAP compared to linear SNAP.
It is important to note that these comparisons involved the same SNAP descriptors for $j_{max} = 4$; we did not optimize SNAP hyperparameters for linear models as was successfully done for other complicated multi-element systems \cite{sikorski2022machine}.
If we chose this route for linear SNAP, it may be possible to include metrics like RDF or MSD as objective functions, and use multi-objective optimization of SNAP hyperparameters to find linear fits that perform as well as NNs.
Such a procedure, however, was not required for NN SNAP.
Overall for LGPS, NN SNAP exhibits better agreement with  AIMD Li ion diffusivity compared to linear SNAP, as well as a nearly two-fold improvement in force errors.

For LiPS, we use the data set available in literature which consists of 520 K AIMD trajectories \cite{batzner20223}.
We were unable properly stabilize LiPS using linear SNAP with this data set; for LGPS we required high temperature AIMD data up to 1600 K for stabilizing linear SNAP.
This is seen by the disagreement of RDFs in Figure \ref{fig:lgps_lips}c.
NN SNAP, on the other hand, does not require high temperature configurations to stabilize this system; the existing LiPS data set in literature was therefore sufficient to accurately model the structure of Li ions.
In Figure \ref{fig:lgps_lips}d we see the improvement offered by using NN SNAP with 56 descriptors ($j_{max} = 4$) compared to 31 descriptors ($j_{max} = 3$).
The best NN SNAP potential exhibits much closer agreement to the AIMD diffusion curve compared to linear SNAP, although not as well as the state-of-the-art graph based NequIP model.
This is not suprising considering the extremely impressive force MAE of 4.7 meV/$\angstrom$ achieved by NequIP for LiPS \cite{batzner20223}.
Our best NN SNAP model obtained a force MAE of ~80 meV/$\angstrom$ with computational cost of $\sim$ $1 \times 10^{-4}$ s per timestep per atom per CPU core.

\begin{figure}[htbp]
\centering
\includegraphics[width=\textwidth]{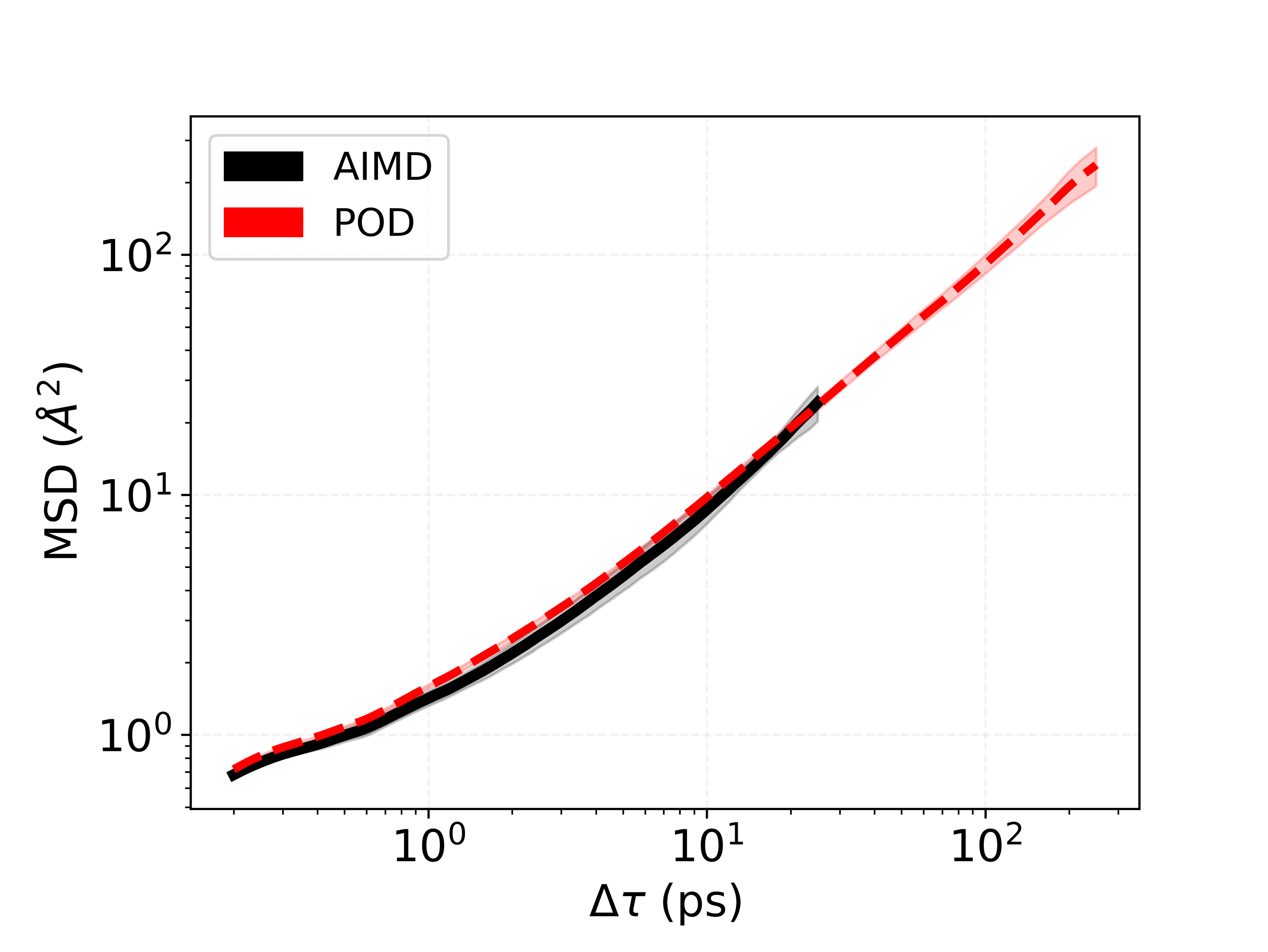}
\caption{Linear POD diffusivity simulations for LGPS at 600 K. POD was trained on the same LGPS dataset as NN-SNAP. This shows that a complex linear model with a sufficient number fitting coefficients can achieve similar AIMD stability and diffusion agreement compared to more flexible NN models.}
\label{msd_pod_allegro}	
\end{figure}

Figure \ref{fig:lgps_lips} might suggest that nonlinear models are simply better or necessary for modelling mass transfer in complex multi-element systems, but this may not be the case. 
A linear basis with sufficiently detailed descriptors can also achieve the same level of AIMD agreement. 
We demonstrate this for LGPS using the recently developed POD descriptors with a linear model. 
LGPS Li ion diffusion simulations using POD \cite{nguyen2023proper} are shown in Figure \ref{msd_pod_allegro}.
As seen in Figure \ref{msd_pod_allegro}, linear POD exhibits excellent agreement with AIMD diffusion like NN SNAP.
Our results here show, for the first time, the ability of linear models to accurately model diffusion in Li superionic conductors.
Linear models possess pros and cons compared to deep learning methods, so this is a worthy addition to the molecular modelling toolkit used by researchers.
If more accuracy is ever desired, our results with SNAP descriptors show that simply feeding these features into a NN can result in a two-fold improvement in force accuracy.
The resulting impact on simulated properties is positive, evidenced by our investigation of Si, GaN, LGPS, and LiPS.

\section{Discussion}

We observed improvement in simulated properties calculated with nonlinear models that use the same features as linear models.
This improvement is attributed to the enhanced force accuracy offered by more flexible nonlinear models compared to linear models.
A fair comparison to quantify this force improvement was achieved by varying loss function force/energy weight hyperparameters, where it was found that all models saturate at a different agreement in forces.
This observation of a force saturation level may be used by researchers to fairly determine which models best match the shape of the PES, irrespective of loss function energy/force weight hyperparameters.
Our observation on the correlation between force accuracy and simulated property accuracy can also aid researchers in determining a force threshold required for their particular application; this is important if researchers want to choose the simplest or cheapest model for the task at hand.

The correlation between force accuracy and simulated property accuracy is best observed with properties that are less prone to cancellation of errors, such as vibrational frequencies.
Higher order properties that are measured by a thermodynamic sampling of states such as thermal conductivity, mass diffusivity, and liquid structure, on the other hand, are known to exhibit cancellation of errors where potentials can achieve poor force/energy errors but still produce accurate property values \cite{rohskopf2020fast}.
It is also important to note that force/energy accuracy is not always enough for obtaining usable potentials that perform stable MD simulations and property prediction \cite{fu2022forces}.
For our data sets and models, however, we showed that the improvement in force errors when using NNs does not sacrifice MD stability.
We observe here that our potentials with most accurate forces exhibit the best property agreement.
This was achieved with Si, GaN, and LGPS by simply feeding descriptors traditionally used with linear models, such as SNAP and ACE, into quadratic or NN models.

Despite the improved accuracy from using NNs with a given set of atom-centered descriptors, it is important to note that linear models with a sufficient basis and larger number of fitting coefficients can outperform NNs with a less descriptive basis or fewer features.
We showed this using the more complicated Li ion systems.
For LGPS and LiPS, an original set of 56 SNAP descriptors input to a linear model was not sufficient to accurately model diffusion.
With the smaller diversity in training data for the LiPS set \cite{batzner20223}, we even had difficulty stabilizing the linear model for MD simulations.
Using this original set of 56 SNAP descriptors with a NN model greatly improved stability along with the simulated mass diffusivity.
This does not mean NNs are required to obtain such accuracies, however, as we illustrated with the recently developed POD descriptors \cite{nguyen2023proper}.
Our POD potential involves a linear model that uses a total of 91 descriptors and 3314 fitting coefficients; this resulted in better force errors than NN-SNAP with 56 descriptors input to a 64 $\times$ 32 NN architecture, and similar agreements in mass diffusivity.
With the systems we studied here, we therefore cannot claim that particular materials or properties always require nonlinear model forms for reliable simulations.
For simpler/smaller sets of descriptors, however, nonlinear model forms are required to achieve the desired level of \textit{ab initio} property agreement showed in this study (1\% error in phonon frequencies, 0.01 MAE in RDF, and $<$ 10 \% error in thermal conductivity and mass diffusivity).
Nonetheless, the results herein show that a linear model with a sufficient basis and number of fitting coefficients can achieve similar results compared to NNs.
It remains to be seen in future work if this will always be the case for systems that are more complicated than the four element LGPS system studied here, such as rare earth metals \cite{hachiya1999interatomic}, metal organic frameworks \cite{bucior2019energy}, or many-element alloys \cite{george2019high}.

Aside from more complicated chemical systems, NNs might also possess an advantage in more complicated simulated properties.
Indeed, our study is limited to properties at equilibrium such as liquid structure, phonon frequencies, and transport properties.
All of these properties require small extrapolations from an equilibrium structure.
It is therefore paramount that researchers heavily weight the equilibrium structure when making potentials for equilibrium properties.
For example, calculating phonon frequencies on a structure with non-zero forces will result in negative frequencies.
This can be alleviated by assigning larger weights in the loss function of Equation \ref{loss_function} for the equilibrium configurations, or oversampling the equilibrium configurations when training.
For non-equilibrium phenomena, however, one may need to sufficiently model multiple equilibrium states as well as transitions between these states.
In such non-equilibrium scenarios, active learning approaches may also be necessary to gather appropriate data, instead of fitting purely to AIMD simulations at equilibrium like we did here.
This is currently a topic of work applied to NN and linear models \cite{zhu2022fast, lysogorskiy2023active}.

\section{Conclusion and Outlook}

Overall we sought to answer the original question of how much gain in accuracy can be expected by feeding descriptors into different ML models, where we saw up to a 50\% improvement in force accuracy when using nonlinear models such as NNs compared to linear models.
For the Si, GaN, and Li ion systems studied here, this resulted in significant improvement in simulated property errors with respect to AIMD simulations.
We can therefore scale up \textit{ab initio} accuracy to significantly larger length and time scales by simply taking existing and widely used descriptors and feeding them into nonlinear models.
Despite this improvement when using nonlinear models with SNAP and ACE descriptors, we also showed that linear models with a sufficient basis can achieve property simulations on par with NNs.
We showed this using linear POD with more fitting coefficients than our NN models.
This is an important result because linear and NN regression both have unique advantages and disadvantages, and molecular modelling researchers may use both of these capabilities with the publicly available tools developed herein.

A fruitful topic for future work is to investigate whether simpler descriptors (e.g. 56 SNAP descriptors) input to complicated models (e.g. NN) possess advantages/disadvantages compared to complex descriptors (e.g. thousands of POD or ACE descriptors) input to simple models (e.g. linear).
There may be other advantages and disadvantages of linear and NN models not studied in detail here, such as MD stability/usability or extrapolation ability in non-equilibrium events such as chemical reactions.
The ability to use both linear and nonlinear models expands the community toolbox for such future studies, and for creating potentials that describe a variety of systems/scenarios.
To this end, we created FitSNAP as an open-source software possessing all the abilities shown in this manuscript; fitting linear and NN models with different descriptors such as SNAP or ACE, then immediately deploying the model for high performance MD in LAMMPS \cite{thompson2022lammps}.
Our linear POD potential is also available as a LAMMPS package where users can perform linear regression with training data and immediately use the potential after.
To conclude, we presented a variety of potentials available for researchers in the open-source LAMMPS/FitSNAP ecosystem and benchmarked expected improvements in fitting and simulation accuracy from using such models for simulating transport properties.

\section*{Author Contributions}
A.R. developed software to support ML models in FitSNAP and LAMMPS, along with curating data sets and property simulations. J.G. implemented ACE descriptors in FitSNAP. D.S. further aided in training and development. C.N. trained POD potentials and aided with the LAMMPS implementation in the ML-POD package. K.G. and  A.H. helped curate LGPS training data and mass diffusion calculations. A.P. and M.W. led the development of FitSNAP and LAMMPS to support new features seen in this study.

\section*{Methods}

All models with SNAP and ACE descriptors were trained using the FitSNAP software, which we provide open-sourced with full documentation on linear and NN training procedures.
The POD potential was trained using least squares on a linear system of equations, as implemented in the LAMMPS ML-POD package.
NN models were trained using a modified form of iterated backpropagation \cite{smith2020simple}.
The iterated backpropagation algorithm we implemented in FitSNAP is as follows.
\begin{enumerate}
\item Calculate all descriptors and their spatial derivatives in LAMMPS.
\item Perform a forward pass that builds a computational graph in an automatic differentation framework; we use PyTorch \cite{paszke2019pytorch}.
\item Perform a backward pass to obtain the derivatives of the NN output with respect to inputs i.e. the array of values $\beta_{jk}$ defined in Eq.~\ref{betas}.
\item Using the array $\beta_{jk}$, calculate atomic forces in LAMMPS with Eq.~\ref{force_general}
\item Perform a second forward pass in PyTorch to calculate loss function defined in Eq.~\ref{loss_function}
\item Perform a second backward pass to obtain loss function derivatives with respect to model coefficients, which are used in gradient descent minimization.
\end{enumerate}

A key aspect of this iterated backpropagation method is that it eliminates the need to store gradients of model outputs with respect to model coefficients for the entire batch.
Instead, we only need to store loss function derivatives with respect to model coefficients.
Previous implementations of force training were strongly limited by available physical memory \cite{Singraber2019}.
Relaxing this constraint makes it possible to explore more diverse combinations of model complexity and force training protocols, such as larger models and/or batch sizes.

\section*{Acknowledgements}
This paper describes objective technical results and analysis. Any subjective views or opinions that might be expressed in the paper do not necessarily represent the views of the U.S. Department of Energy or the United States Government. A.R., J.G., A.T., and M.W. acknowledge funding support from the Exascale Computing Project (17-SC-20-SC), a collaborative effort of the U.S. Department of Energy Office of Science and the National Nuclear Security Administration, and U.S. Department of Energy, Office of Fusion Energy Sciences (OFES) under Field Work Proposal Number 20-023149.
Dr. Nguyen and Dionysios Sema acknowledge the United States  Department of Energy under contract DE-NA0003965.
Dr. Nguyen also acknowledges the Air Force Office of Scientific Research under Grant No. FA9550-22-1-0356 for supporting his work.
K.G. and A.H. acknowledge support from the National Science Foundation (NSF) career award to A.H. (award no. 1554050) and the Office of Naval Research (ONR) under a MURI program (grant no. N00014-18-1-2429).

This article has been authored by an employee of National Technology and Engineering Solutions of Sandia, LLC under Contract No. DE-NA0003525 with the U.S. Department of Energy (DOE). The employee owns all right, title and interest in and to the article and is solely responsible for its contents.
The United States Government retains and the publisher, by accepting the article for publication, acknowledges that the United States Government retains a non-exclusive, paid-up, irrevocable, world-wide license to publish or reproduce the published form of this article or allow others to do so, for United States Government purposes. The DOE will provide public access to these results of federally sponsored research in accordance with the DOE Public Access Plan https://www.energy.gov/downloads/doe-public-access-plan.

\section*{Declaration of Competing Interests}

The authors declare that they have no known competing financial interests or personal relationships that could have
appeared to influence the work reported in this paper.

\section*{Data Availability}

Training data, FitSNAP input scripts, and LAMMPS input scripts for mass diffusion simulations are available on our FitSNAP GitHub page.


\bibliography{sn-article.bib}

\end{document}